\documentclass[11pt, onecolumn, draftcls ]{IEEEtran}
\usepackage{cite}
\usepackage{algorithm2e}
\usepackage{amssymb}
\usepackage{enumerate}
\usepackage{graphicx}
\usepackage{amssymb}
\usepackage{amsbsy}
\usepackage{float}
\usepackage{balance}
\usepackage{subfigure}
\usepackage{url}
\usepackage{bm}
\newfloat{longequation}{b}{ext}
\newfloat{longequation}{t}{ext}

\usepackage[cmex10]{amsmath}
\usepackage{array}
\usepackage{varioref}
\usepackage[normalem]{ulem}
\graphicspath{{./figs/}}
\usepackage{algorithm2e}
\usepackage{amssymb}
\usepackage{graphicx}
\usepackage{amssymb}
\usepackage{amsbsy}
\usepackage{float}
\usepackage{balance}
\usepackage{subfigure}
\usepackage{url}
\usepackage{bm}
\usepackage[cmex10]{amsmath}
\usepackage{array}
\usepackage{varioref}
\usepackage[normalem]{ulem}
\graphicspath{{./figs/}}
\usepackage{cite}
\usepackage{color}

\begin{document}

\title{Coordinated  Multibeam Satellite Co-location:\\ The Dual Satellite Paradigm}
\author{ Dimitrios Christopoulos$^\star$, Shree Krishna Sharma$^\star$, Symeon Chatzinotas$^\star$, Jens Krause$^\dag$ and Bj$\ddot{\mathrm{o}}$rn Ottersten$^\star$ \\
$^\star$Interdisciplinary Centre for Security, Reliability and Trust (SnT), University of Luxembourg \\
Email: \{shree.sharma, dimitrios.christopoulos, symeon.chatzinotas, bjorn.ottersten\}@uni.lu.\\
$^\dag$ SES, Chateau de Betzdorf, Luxembourg,
Email: jens.krause@ses.com
}

\maketitle


\begin{abstract}
    In the present article, a new system architecture for the next generation of satellite communication (SatComs) is presented. The key concept lies in the collaboration between multibeam satellites that share one orbital position. Multi-satellite constellations in unique orbital slots offer gradual deployment to cover unpredictable traffic patterns and redundancy to hardware failure advantages. They are also of high relevance  during the satellite replacement phases or necessitated by constraints in the maximum communications payload that a single satellite can bear. In this context,  the potential gains of advanced architectures, that is architectures enabled by the general class of cooperative and cognitive techniques, are exhibited via  a simple paradigm.  More specifically, the scenario presented herein,  involves two co-existing multibeam satellites  which illuminate overlapping coverage areas. Based on this scenario, specific types of cooperative and cognitive techniques are herein considered as candidate technologies that can boost the performance of multibeam satellite constellations. These techniques are compared to conventional frequency splitting configurations in terms of three different criteria, namely the spectral efficiency, the power efficiency and  the fairness.  Consequently,  insightful guidelines for the design of future high throughput constellations of multibeam satellites are given. 
\end{abstract}
\section{Introduction}
\subsection{Satcoms: State-of-the Art}
Nowadays, the main application of fixed satellite services involves broadcasting information to a large number of user terminals, distributed over a wide coverage. In spite of the market driven, broadcasting nature of current satellite communications (SatComs), the road towards interactive broadband services seems inevitable. Multibeam satellite systems that reuse the available spectrum have already managed to provide broadband services, facilitated by the second generation of satellite standards \cite{DVB_S2_standard}. For instance,  Viasat1 \cite{viasat1} can deliver up to 110~Gbps of total throughput over the coverage. A new generation of multibeam systems that still reuse frequency in a conventional manner is expected by 2016 \cite{viasat1}.  Nevertheless, as the most recent extensions of SatCom standards have shown \cite{DVB_S2X}, there is only as much as one can achieve with fractional frequency reuse and conventional payloads. Therefore, satellite manufactures are  exploring novel  system architectures \cite{vidal2013arch} that can match the expected demand.
\subsection{Multibeam SatComs: beyond SoA.}
Advanced satellite system architectures, able to meet the highly increasing demand for throughput and close the digital divide are of high relevance nowadays. In this direction, the investigation of aggressive frequency reuse methods comes into play. The term aggressive frequency reuse, refers to operating a fractional reuse system, such as a multibeam satellite system,  with very low frequency reuse factors. In the extreme case of full frequency reuse, all available bandwidth is allocated to all beams, leading to a reuse factor of one and a high level of co-channel interferences. Such configurations are enabled by the spatial degrees of freedom offered by the multibeam antenna.     To fully exploit  this spatial separation,  advanced signal processing techniques, namely precoding, constitute a substantial interference mitigation resource \cite{Christopoulos2013AIAA, Zheng2012, Christopoulos2012_EURASIP}. As a result,  the scarce user link bandwidth can be efficiently utilized by higher frequency reuse schemes.
The most recent results on aggressive frequency reuse multibeam satellites can be found in \cite{Christopoulos2014_TWCOM} and the references therein.

 Taking  the above concept a step further,  aggressive frequency reuse can come into play  between physically separated satellites \cite{Christopoulos2012_ICC}. The term dual multibeam satellites, will hereafter refer to satellites bearing multibeam   communications payloads compatible with aggressive frequency reuse configurations, that share one orbital position. Under the assumption of    information exchange between the different gateways (GWs) that serve each satellite, advanced interference mitigation techniques can come into play. The techniques considered in the present work can be classified in the general class of  cooperative and cognitive methods. Prior to introducing these techniques,  the motivation behind dual multibeam satellite scenarios is given.
\subsection{Satellite Co-location}
The herein examined dual satellite paradigm constitutes an instance of a constellation of multiple satellites, co-located in a single orbital slot. Satellite co-location was pioneered by SES with the Astra 19.2$^\circ$E system for the delivery of broadcasting services \cite{HighAbove}.
The reasoning behind this proven concept and its extension to multibeam satellite architectures is summarized by the following points:
 \begin{itemize}
   \item  \textit{Orbital slot Congestion:} In the evolution of {geostationary (GEO)} satellite systems, orbital slots are becoming a scarce resource. To address this uprising problem, the deployment of more than one multibeam satellites, in one orbital position becomes relevant.
   \item \textit{Traffic Demand:} The operational lifetime of a multibeam satellite spans over a period of more than fifteen years. It is therefore probable that unpredictable changes in the traffic demand might dictate the launch of secondary satellites to support existing ones. The opportunity to place such satellites in the same orbital slot, is considered as a great asset for the satellite operator. What is more, and even if traffic demand is well predicted, the gradual deployment offered by a multi-satellite system reduces the upfront investment and the operational cost, thus providing higher flexibility to the operator.
   \item \textit{Payload Complexity: } Aggressive frequency reuse increases the communication payload size since a single high power amplifier (HPA) cannot be shared by multiple beams \cite{Christopoulos2014_TWCOM}. Hence, the payload required to drive  a large number of beams that cover large regions (e.g. pan-European coverage) can  be carried by multiple  co-existing satellites.
    \item \textit{Redundancy:} Hardware redundancy to guarantee uninterrupted service delivery in case of malfunctions is of high importance for SatComs. The co-location of  separate platforms in a single orbital slot reduces the individual payload requirements, thus allowing for  redundant equipment to be carried for the cases of failure. More importantly, redundancy can be offered between the co-located satellites. 
    \item  \textit{A priory co-location:} Last but not least, long periods of coexisting satellites appear by default during the satellite replacement phase. This \textit{a priori} co-location can be exploited towards increasing the system capacity.
\end{itemize}

Based on the above arguments, this work attempts to determine the most promising, with respect to specific performance metrics,  method to enable the cooperation of multibeam satellites, co-located   in a unique orbital slot.
In the following section, the dual satellite paradigm, an instance of a multibeam satellite constellation, is formulated.

\section{Dual Satellite Paradigm}
 \begin{figure}[h] \centering
 \includegraphics[width=0.8\columnwidth]{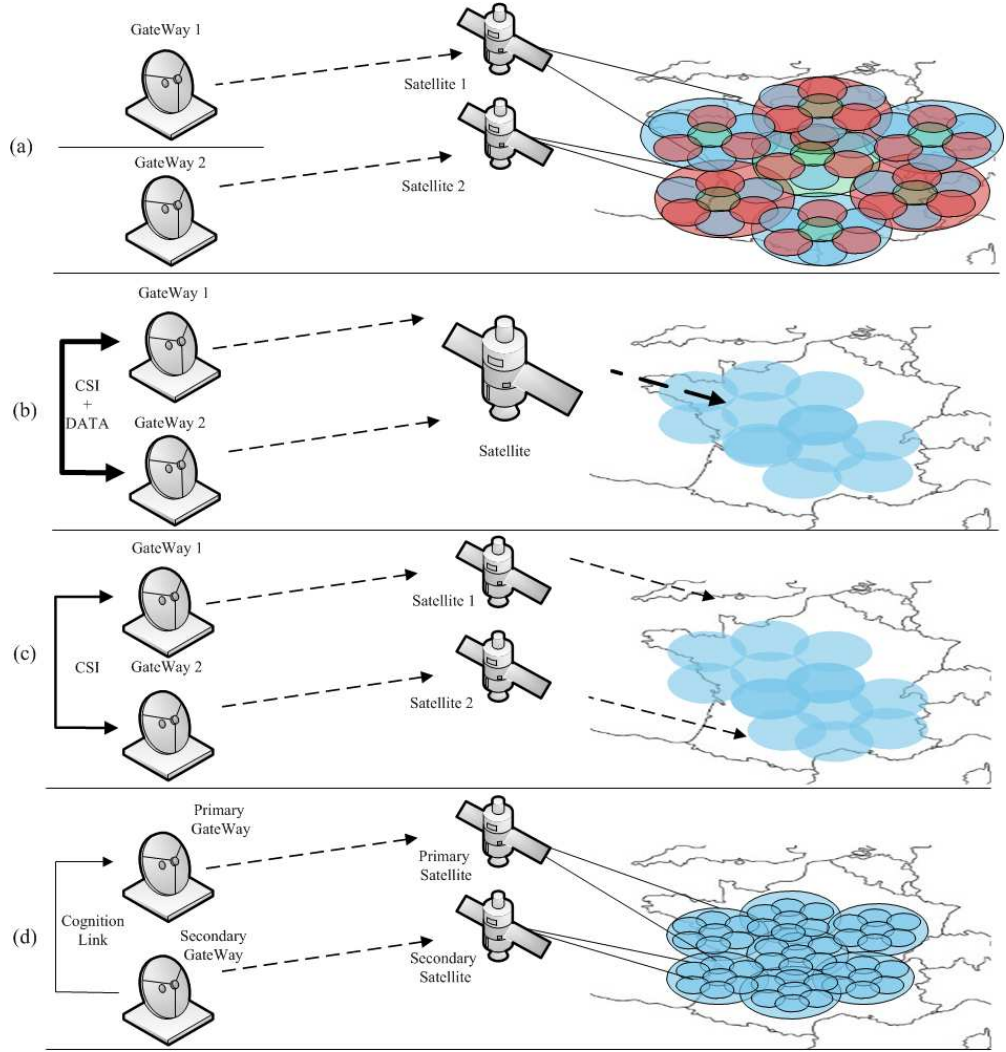}\\
 \caption{Different architectures to realize constellations of co-existing satellites. Different colors represent different frequency bands.}\label{fig: system model}
 \end{figure}

In Fig. \ref{fig: system model}, four possible ways to deploy multibeam satellites in one orbital position are presented. More details for each architecture are given hereafter.
\subsection{Conventional  frequency splitting}
The simplest way to facilitate satellite co-location  requires no added complexity and is illustrated in Fig. \ref{fig: system model} (a). In this scenario, the total available bandwidth of the forward link is divided into two equal segments. Further on, this bandwidth is divided into $N_{\mathrm{c}}$ segments, where the parameter $N_{\mathrm{c}}$ is the frequency reuse factor.  The total gain in terms of frequency reuse obtained by using a multibeam satellite  depends on the frequency reuse factor.  As the value of $N_{\mathrm{c}}$ decreases, the available bandwidth per beam increases, at the expense however, of increased co-channel interference. Remembering the Shannon formula, the capacity scales linearly with the bandwidth and in logarithmic fashion with the signal to interference ratio. However, due to the antenna pattern, multiple tiers of interference are introduced when frequency is aggressively reused. Thus,  the value of $N_{\mathrm{c}}$ should be chosen in such a way that the maximum system capacity is achieved. Herein, a frequency reuse factor of three is considered.

\subsection{Cooperation}
A cooperative dual satellite system refers to two  satellites bearing aggressive frequency multibeam communications payloads, that are fed by fully interconnected and  synchronized on a symbol level GWs. Under these assumptions, advanced signal processing techniques, namely linear precoding,  can be applied and the two transmitters will ideally behave as one large satellite that bears the equivalent payload of the two platforms as depicted in Fig.   \ref{fig: system model} (b).  This fact, greatly increases the available degrees of freedom thus maximizing the potential gains of such systems.
However, the stringent demand of synchronization between two physically separated satellites renders such a scenario highly challenging.
 Despite this, it is herein considered as an upper bound of the presented techniques.
To avoid a highly complex architecture, partial cooperation is proposed in the following.
\subsection{Coordination }
  A simplest approach is  the partial cooperation, hereafter referred to as coordination, between the two coexisting transmitters, as shown in Fig.   \ref{fig: system model} (c). In this manner, the total system performance can be increased while maintaining system complexity at moderate levels. The term coordination implies a relaxation in the synchronization and the data exchange requirements. More specifically,  coordination involves the exchange of a smaller amount of data, namely channel state information (CSI) and does not require the joint processing of signals between the two GWs. Therefore, it  trades-off the high gains of inter-system cooperation for a reduced implementation complexity. After the exchange of CSI, each satellite  serves only a set of users. Hence, the signals transmitted by each satellite do not need to be synchronized on the symbol level. More details on the algorithm to determine the user allocation to each satellite will be provided in the respective sections.
\subsection{Cognition}
Cognitive communications are considered a promising tool  to address the spectrum scarcity problem caused by spectrum segmentation and current static frequency allocation policies \cite{Gridlock}. Several Cognitive Radio (CR) techniques have been proposed in the literature in order to allow the coexistence of cognitive systems with the licensed primary systems. The most common cognitive techniques  can be categorized into interweave or Spectrum Sensing (SS), underlay, overlay and Database (DB) related techniques. In SS only techniques, Secondary Users (SUs) are allowed to transmit whenever Primary Users (PUs) do not use a specific band, whereas in underlay techniques, SUs are allowed to transmit as long as they meet the interference constraint of the PUs. Overlay networks are characterized by the mitigation of interference with the help of advanced coding and transmission strategies at the cognitive transmitters while in the DB scenarios, cognitive terminals query the predefined DB in order to find the unoccupied frequency bands and utilize them.
The potential of CR in terrestrial systems has aspired the concept of cognitive SatComs.  In the field of cognitive SatComs, the main  related literature  can be found in \cite{Sks:jsat14} and the references therein. Despite the fact that cognition has also been assessed for the coexistence of satellite and terrestrial systems, the present work will only focus on the cognition between satellite systems. Consequently, possible conflicts of interest between satellite and terrestrial providers over the scarce spectrum are avoided. 

In this article, the technical aspects  coordinated dual satellite systems are over-viewed in Sec. \ref{sec: coordinated}. Next, Sec. \ref{sec: cognitive} presents a simple cognitive techniques that will be considered herein. Following this, the two approaches are numerically compared in Sec. \ref{sec: results}. Finally, the challenges for the application of such architectures are described in Sec. \ref{sec: challenges}.


\begin{table}
\caption{Link Budget and Simulation Parameters}
\centering
\begin{tabular}{l|c}
\textbf{Parameter}  & \textbf{Value}  \\\hline
 Frequency Band  &  Ka (20~GHz)\\
 Total user link Bandwidth    & 500~MHz \\

 Multibeam Antenna Gain   & Bessel Approx. \cite{Christopoulos2012_EURASIP} \\
 Total on-board  Power $P_{\mathrm{tot}}$ & 29~dBW \\
\hline
\textbf{Per beam Link Budget}\\
 Saturated power per beam  & 55~W \\
 OBO& 5~dB \\
 Transmit power per beam  & 17.38~W\\
 3-dB Beam Gain  & 54~dBi\\
 EIRP  &     66 dBW~ \\
Bore sight distance & 37569~Km \\
  Path Loss & 210 dB\\
  UT antenna gain  & 41~dBi\\
  Carrier Power $C$&-103~dBW\\
  Clear Sky Temperature &235.34K\\
  Noise Power $N$ &-118~dBW\\
  $C/N$&15~dB
 \\\hline
\end{tabular}
\label{tab: LinkBudgParas}
\end{table}

\section{Coordinated   Constellations}\label{sec: coordinated}

  The focus of this work will be limited to coordinated dual multibeam satellites that employ linear precoding and user scheduling to  enable dual multibeam co-location. In the multiuser multiple input single output (MU MISO) literature, precoding is an interference precancelation technique that exploits the spatial degrees of freedom offered by the multiple transmit antennas 
to serve multiple single antenna users. Multiuser interferences are canceled by multiplying the transmit signals by precoding vectors. Thus equivalent interference free channels are created by the transmitter. However, full knowledge of the channel is necessary at the transmitter. The focus is on the MU MISO broadcast channel (BC).
   In the full frequency reuse scenario of  Fig. \ref{fig: system model} (c), interferences from the adjacent satellite are limiting the system, while intra-satellite multiuser interferences are completely mitigated by linearly precoding the transmitted signals in satellite. Simple zero forcing (ZF) precoding methods with per-antenna power constraints are considered herein \cite{Christopoulos2012_ICC}.

The coordinated dual satellite concept assumes that the two satellites can serve a joint pool of available users. The CSI of each user and its data is readily available to both GWs serving each satellite. However, based on the CSI, in each satellite a different set of users is allocated. Therefore, at each transmission instance, each user is served by only one multibeam satellite. This assumption relaxes the necessity to jointly process signals in both GWs. More importantly, it relaxes the constraint of a symbol based synchronization between the two satellites.
\subsection{User Selection and Allocation}\label{sec: SIUA}

As proven in \cite{Yoo2006b}, user selection can significantly improve the performance of ZF in  an individual system. Therefore, the performance of each satellite separately is optimized by constructing a semi-orthogonal user group from a pool of users according to the Semi-orthogonal User Selection (SUS) algorithm of \cite{Yoo2006b}.
However, considering the coexistence of two separate transmitters, as is the case in a dual satellite system,  the problem of high intersatellite interferences arises.
 The main constraint is that the exact calculation of the level of interferences in each iteration is not possible since the exact user set is still undetermined. However, based on a basic advantage of ZF beamforming, which is the decoupled nature of the precoder design and the power allocation optimization problems, an approximation of the interferences can be made\cite{Christopoulos2012_ICC}. To the end of managing the inter-satellite interferences,  a novel algorithm that selects users and allocates them  to each satellite has been proposed in \cite{Christopoulos2012_ICC}. This  algorithm accounts for the effects of  the interferences between the two satellites, and is hereafter described.

The Semi-orthogonal Interference aware User Allocation (SIUA) algorithm of  \cite{Christopoulos2012_ICC}, improves the performance of each satellite and of  the overall system, simultaneously. The first is achieved by maximizing the orthogonality between users allocated in the same set, hence optimizing the ZF performance,  whilst the second by minimizing the level of interferences between the two sets. Consequently, the SIUA algorithm  finds the most orthogonal users that at the same time receive and induce the least possible interferences. This algorithm, requires knowledge of the CSI of all users. Therefore,   each GW handles only the data of the users allocated to the corresponding satellite and thus the amount of data that needs to be exchanged is reduced.

\section{Cognitive  Constellations}\label{sec: cognitive}
Although the cognitive radio literature is quite mature in the terrestrial context, the application of cognition in SatCom systems is still in its infancy. Satellite systems operating in same or different orbits can be employed to provide different satellite services over the overlapping coverage area using the same frequency resources. One satellite system can be assumed to be primary and to have priority over the shared spectrum while another satellite system operates in a secondary way by providing sufficient protection to the existing licensed users. Several dual satellite co-location scenarios may exist in future SatCom networks and can be categorized on the basis of (i) operating frequency, (ii) operating mode, (iii) operators' ownership, (iv) coverage type, and (v) satellite orbit. Depending on the scenario, several techniques such as cognitive interference alignment, spectrum sensing, cognitive beam-hopping, power control, exclusion zone, etc.  have been identified in the SatComs related literature  \cite{Sharma2013VTC}. Herein, the focus  will be limited to cognitive beam-hopping. For the cognitive architecture of Fig. 1 (d),
the primary satellite generates large beams over the coverage while the
secondary deploys smaller ones over the same coverage area.
\subsection{Cognitive Beamhopping System}
The term beamhopping refers to a system in which a portion of the total beams are simultaneously active with a regular repetition pattern\cite{Fso:12}.   This technique applies a regular time window periodically  and thus allows for the entire available bandwidth to be allocated to each illuminated beam. The duration for each illuminated beam should be selected to satisfy the user transmission delay requirement.
In more detail, the cognitive beamhopping system  originally proposed in \cite{Sks:jsat14} is herein considered.   Based on this a priori knowledge of the beamhopping pattern, the secondary satellite's beamhopping pattern is designed so that it does not degrade the primary's operation. The primary system employs the slot reuse factor of three and the secondary satellite adjusts its beamhopping pattern to avoid the primary active beams. The primary and secondary transmissions can be synchronized with the help of the  timing information the primary satellite provides.
Moreover, the cognitive beamhopping  with the power allocation method proposed in \cite{Sks:jsat14} is also considered. 

The comparison of the techniques to determine the optimal in a system throughput, fairness and energy efficiency sense,  technique follows.

\section{Performance Comparison} \label{sec: results}

A baseband block fading model, as described in \cite{Christopoulos2012_EURASIP}, models the satellite antenna radiation pattern, the path loss, the receive antenna gain and  the noise power.
Clear sky conditions are assumed.
The users, each equipped with a single antenna, are  uniformly distributed over the coverage area.  Only one user per beam is served in each transmission.  Despite the fact that in each satellite separately,  a MU MISO BC     is realized, the total system operates over an interference channel.
The link budget considerations  are included in Tab. \ref{tab: LinkBudgParas}, along with an instance of nominal link budget. This corresponds to the on-board available power of current operational multibeam satellites and is included  to provide a point of reference. However, to investigate the trends of the proposed methods with respect to the transmit power, results are plotted in a range of on board available power. 
\subsection{Spectral Efficiency}
\begin{figure}[h] \centering
  \includegraphics[width=0.9\columnwidth]{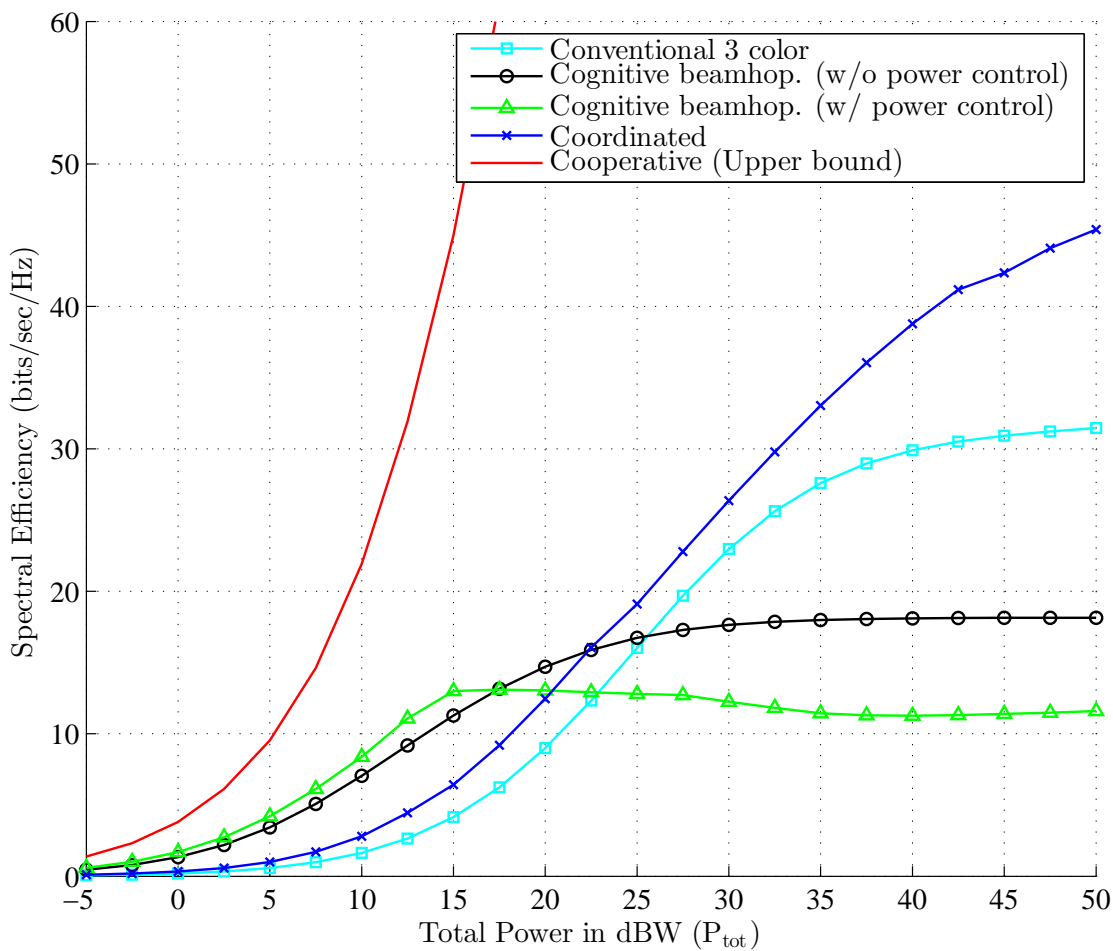}
  \caption{Spectral efficiency of the proposed schemes}\label{fig: spf}
  \end{figure}

In the present section, the  performance evaluation and comparison  of cooperative and cognitive dual satellite systems is performed in terms of spectral efficiency (bits/sec/Hz). For a proper comparison,  the same total power budget is employed in the different architectures.
Figure \ref{fig: spf} presents the Spectral Efficiency (SE) versus the total power budget $P_{\mathrm{tot} } = [-5: 50] $~dBWs. In the nominal operation point, i.e. 29 dBWs, the gain of the coordinated system over other approaches is notable. Also, as  the available power increases, further gains can be gleaned by the coordinated systems. These gains stem from the saturation of the SE performance of conventional designs due to interbeam interferences. Despite the fact that the herein considered cognitive beamhopping techniques are greatly affected by interferences, their value is noted in the low power, noise limited regime.  Cognitive beamhopping with power control becomes also relevant for very low power budgets, while in the mid ranges, power control is not offering any gains.  Further, as it is illustrated in Fig. \ref{fig: spf},  there exists a crossing point between the performance curves of the coordinated dual satellite  and cognitive beamhopping (without power control) systems  at the value of $P_{\mathrm{tot}}=22.5$ dBW. Also, at this value, all considered methods perform no worst than the conventional systems. Consequently, this value will serve herein as a threshold for the choice of the most beneficial from a SE perspective, technique. as well as a point at which the techniques can be compared with respect to other criteria.  Finally, The MU MIMO channel capacity curve is also plotted in Fig. \ref{fig: spf}. Clearly, it can be noted that in the  lower power region, the performance gap from the channel capacity of the cognitive methods  is reduced. However, coordination manages to maintain a smaller to the theoretical upper bound gap than the other techniques, in the high power region.
\subsection{Instantaneous Fairness}
\begin{table}
\caption{Instantaneous Fairness Comparison}
\centering
\begin{tabular}{l|c|c}
\textbf{Technique}  &\textbf{Fig. 1}& \textbf{Jain's Index} \cite{Hua:14}  \\\hline
Conventional 3 Color& (a)& $0.766$\\
Coordinated &(c)& $0.127$\\
Cognitive (w/o power control)&(d)&$0.254$\\
Cognitive (w/ power control)&(d)&0.201\\\hline
\end{tabular}
\end{table}
Increased skepticism over spectrally efficient multibeam satellites stems from the effects of such configurations on the fairness of the system. The goal of the methods considered herein, is to increase the total throughput via aggressive frequency reuse. The present section aims at capturing and quantifying the effects of the proposed methods on the instantaneous fairness. However, it should be stress that the long term, average fairness can be guaranteed by proper user scheduling, which will remain out of the scope of the present work.
 
 Amongst various methods  to depict whether the  rates are equally distributed over the users   in wireless networks \cite{Hua:14}, herein, we apply Jain's fairness index   \cite{Sks:jsat14}. When this index is equal to one, then all users are treated equally. The Jain's index for all approaches and for a total power budget of $P_{\mathrm{tot}} =22.5$ dBW is given in Tab. II.   This is the  point  where the proposed methods perform equally or better than the conventional in terms of SE  (cf.  Fig. \ref{fig: spf}). Intuitively, the fairness reduction of the proposed systems is expected, based on the well known fairness versus sum rate tradeoff in multiuser systems. Clearly, the proposed methods greatly reduce the system fairness, as seen in Tab. II. This is the price paid for more than 30\% of SE gains over conventional approaches. Between the proposed methods, however,  the highest fairness is achieved by  cognitive beamhopping without power control,  since for the same system sum rate,  a fairness index of more than 0.25 is attained. By introducing power control, a  5\% reduction in the fairness index is noted.

\begin{figure}[h] \centering
  \includegraphics[width=0.9\columnwidth]{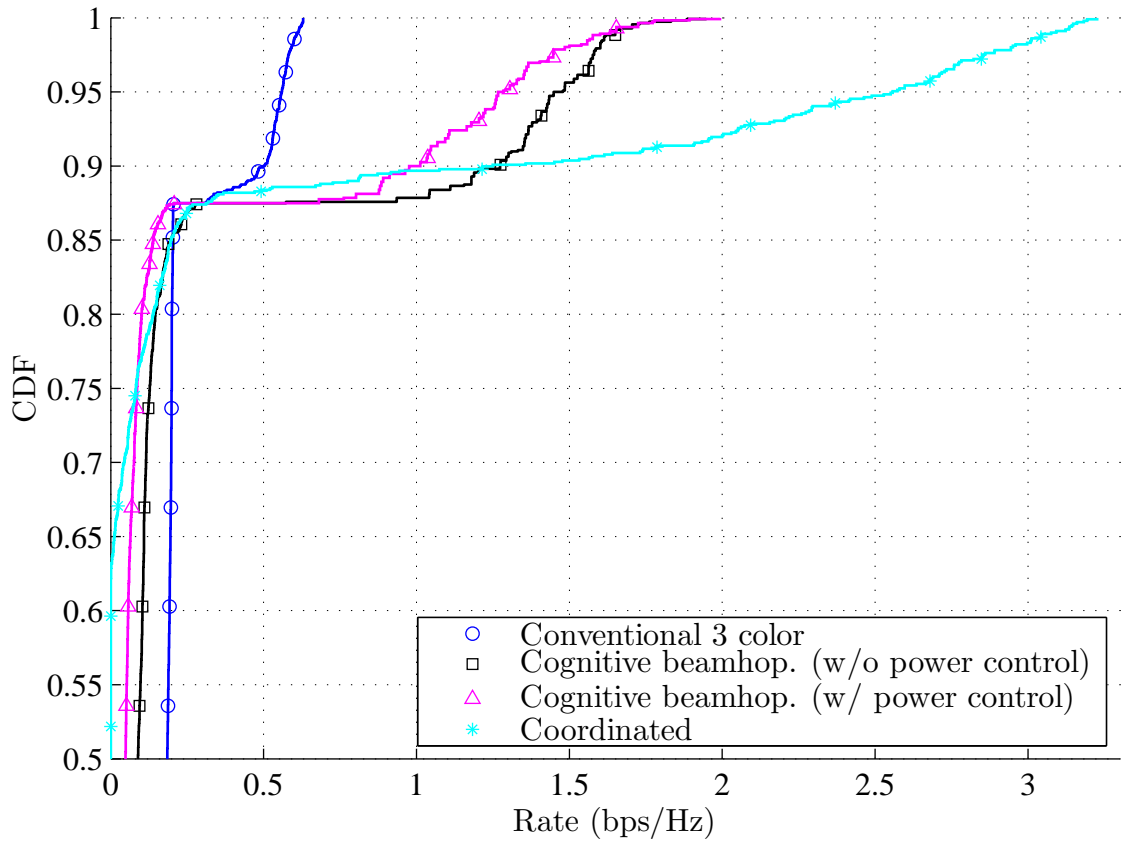}
  \caption{Spectral Efficiency (SE) distribution over the coverage for the proposed methods}\label{fig: CDF}
  \end{figure}

 Moreover, to provide a more concrete  illustration of the fairness criterion,  Fig. \ref{fig: CDF} presents the Cumulative Distribution Function (CDF) curves of the per user SEs provided by each of the considered  schemes again at the value of $P_{\mathrm{tot}} =22.5$ dBW.   In this figure,  the very low rate variance of conventional systems is again clear. Also,  cognitive beamhopping without power control provides better user fairness than the coordinated methods.
Nevertheless, the coordinated dual satellite system achieves more than double rates, at the expense of driving almost 65\% of the users to the unavailability region. At this point, it should be stressed, that the proposed coordinated system is focused on delivering high throughput. Different linear precoding methods than  maximizing the fairness criterion remain out of the scope of this work.
\subsection{Power Efficiency}
\begin{figure}[h] \centering
  \includegraphics[width=0.9\columnwidth]{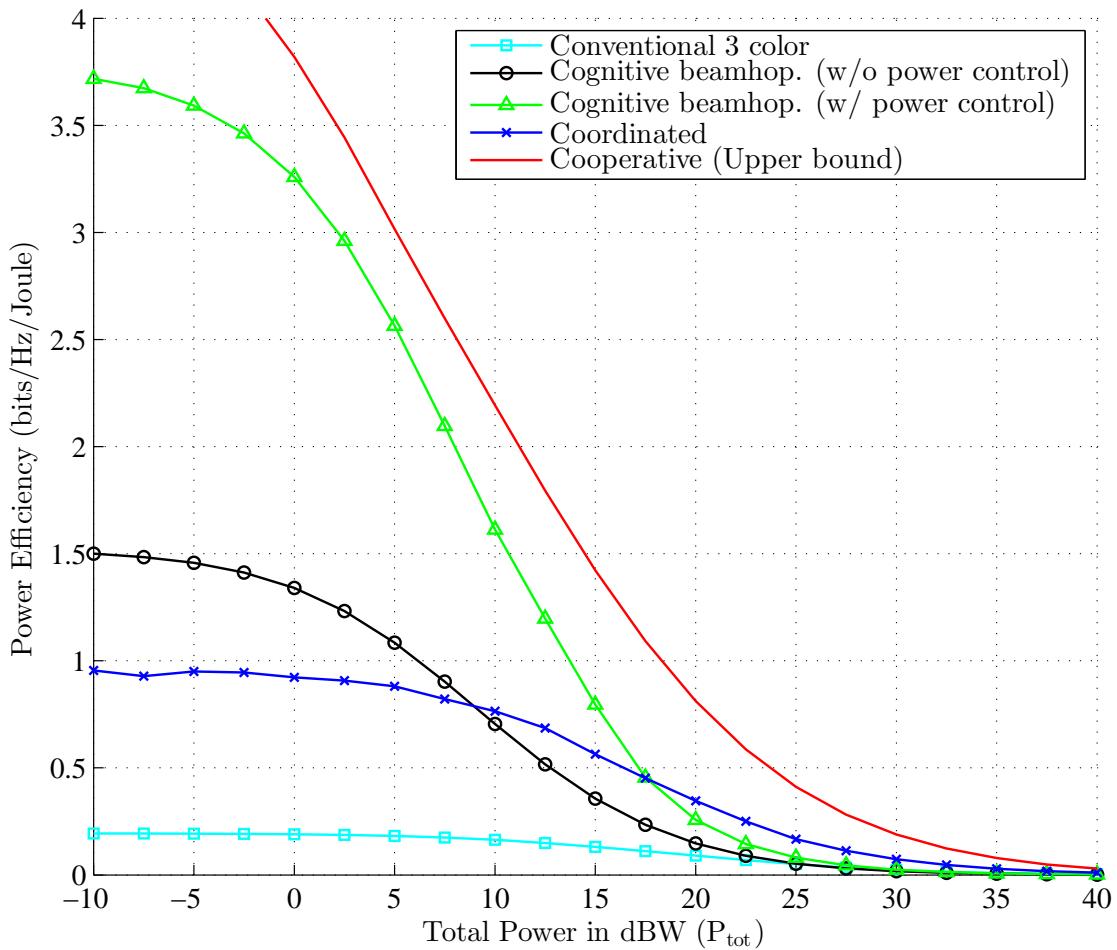}
  \caption{Power efficiency of the proposed schemes}\label{fig: PE}
  \end{figure}

Finally, in Fig. \ref{fig: PE}, the Power Efficiency (PE) versus the total power budget for the considered schemes is plotted. The power efficiency for each scheme is obtained by dividing the SE (cf. Fig. \ref{fig: spf}) by the total amount of the consumed power. From this figure, it is clear how the power efficiency of the  coordinated dual satellite system is better than all other schemes at higher values of $P_{\mathrm{tot}}$, i.e. above $P_{\mathrm{tot}}=17$ dBW. In the lower power regime, cognitive beamhopping with power control manages to outperform all other realizable schemes. Actually, it is  almost as efficient as the optimal cooperative system that serves as an upper bound. This result renders the cognitive beamhopping scheme with power control as the most promising approach for low rate,  power efficient designs. This intuitive concept manages to approach the upper bound of the optimal system in terms of efficient utilization of the power resources at very low implementation costs. If the efficiency and throughput performance is however required, one has to adhere to the more complex coordinated architectures and sufficient power budgets.

 \section{Challenges and Way Forward}\label{sec: challenges}
 \subsection{Performance vs Complexity Tradeoffs }
Despite the fact that the technology readiness level (TRL) of aggressive frequency reuse configurations,  is considered high, especially with the introduction of low mass HPAs, the important issue to enable the methods considered herein is the CSI that needs to be readily available at all GWs. Channel acquisition and synchronization methods, based on the recommendations of the latest DVB guidelines (see Annex E of \cite{DVB_S2X})  are part of future work. In terms of synchronization, coordination relaxes the constraint of symbol level synchronization between the two physically separated satellites, during transmission. However, user scheduling needs to be performed in a joint an synchronized manner. The proposed  SIUA, can be executed in a centralized location or run in parallel at the GWs that share CSI. 

In the  intuitive, cognitive beamhopping scenario,  the payload complexity remains low and no signal processing is required at each satellite either. Since a subset of the total beams is simultaneously active, the payload requirements in terms of HPAs are less. Thus, such a technique helps to reduce the number of amplifiers on board as well as the power demands on the payloads.  However, advanced switching multiplexers are required to support the beamhopping operations. Also,  the multibeam pattern of the secondary satellite is much denser, which implies an added payload complexity. The size of each smaller beam is also  limited by wave diffraction rules.  In terms of GW interconnection, the cognition is achieved by sharing the beamhopping pattern and the timing information of the primary satellite to the secondary system. This is achieved by a signalling link from  the primary GW  towards the secondary. Hence, the connectivity  requirements between the GWs are less compared to the coordinated case,  since a one directional link needs to be implemented (cf. Figs. 1 (c) and (d)).

Based on this complexity discussion, the increased performance of the more complex coordinated techniques is justified. This complexity versus performance tradeoff proves that cooperative and cognitive techniques are complementary to each other and constitute a substantial tool for the design of future SatComs.

 \subsection{Main challenges in realizing dual satellite systems}

A helpful step towards the realization of the   innovative satellite system architectures, is their acceptance from standardization bodies. Despite the fact that coordinated and cognitive satellite constellations are not mature in terms of TRL, the road to their standardization in the next generation of DVB satellite standards needs to be predefined. Following the example of advanced interference mitigation techniques for SatComs\cite{Christopoulos2014_TWCOM}, which have been included in \cite{DVB_S2X},  several practical constraints first need to be incorporated. By establishing practical ways to solve the framing constraints, the channel acquisition problem as well as other inherent satellite channel impairments (e.g. non-linearities of the on-board amplifiers), the acceptance by the  standardization bodies will be facilitated. In the same manner, although cognitive satellite  related standards are also being realized, e.g.  \cite{ETSI_COG}, the satellite co-location scenario has  yet to be considered.   Still, by accounting the potentially high gains at considerably low costs, especially when multibeam satellite constellations will be readily available, this road appears to be worth following.

  \section{Conclusions}
 In the present article, a simple paradigm that substantiates the deployment of cooperative and cognitive architectures in the next generation of satellite communications is given.  On the basis of the dual satellite scenario, when the design aims to enhance the overall system throughput,   the coordinated and cognitive schemes can be alternated, based on the available on-board power. As a rule of the thumb, if the power budget is sufficient, then interference limits the system and thus coordinated systems are the way forward. On the other hand, in power limited systems, the cognitive approaches should be preferred. In terms of power efficiency similar results are noted: coordinated systems are better in the interference limited regime while cognitive are the choice for  noise limited systems. Finally, the reported gains come at the cost of reduced instantaneous fairness. Based on the included arguments  the potential of multibeam satellite co-location by the means of cooperation and cognition, motivates further examination and research on this topic.
\section*{Acknowledgments \& Related Activities}

This work was   partially supported by the National Research Fund, Luxembourg under the project  ``$CO^{2}SAT$'' and $\mathrm{SEMIGOD}$. A related activity is ``$\mathrm{CoRaSat}$: Cognitive Radio for Satellite Communications'', funded by the European framework program. The views expressed herein, can in no way be taken to reflect the official opinion of SES.
\bibliographystyle{IEEEtran}
\bibliography{refs/IEEEabrv,refs/conferences,refs/journals,refs/books,refs/references,refs/csi,refs/thesis}
\end{document}